\documentclass[fdp,a4paper,fleqn%
]{w-art}
\usepackage{times,cite,amsmath,amssymb}

\usepackage[]{graphicx} 

\begin{document}
\DOIsuffix{theDOIsuffix}
\Volume{55}
\Month{01}
\Year{2007}
\pagespan{1}{}
\keywords{gauge/gravity duality, superstring theory.}



\title[Does Yang-Mills theory describe {\it quantum} gravity?]{Does Yang-Mills theory describe {\it quantum} gravity?}


\author[M. Hanada]{Masanori Hanada\inst{1,2,}%
  \footnote{
  Also at Stanford Institute for Theoretical Physics,
Stanford University, Stanford, CA 94305, USA. \\
  E-mail:~\textsf{hanada@yukawa.kyoto-u.ac.jp}}}
\address[\inst{1}]{Yukawa Institute for Theoretical Physics, Kyoto University,
Kitashirakawa Oiwakecho, Sakyo-ku, Kyoto 606-8502, Japan}
\address[\inst{2}]{The Hakubi Center for Advanced Research, Kyoto University,
Yoshida Ushinomiyacho, Sakyo-ku, Kyoto 606-8501, Japan}
\begin{abstract}
The strongest version of the gauge/gravity duality conjecture relates the $1/N$ correction in super Yang-Mills theory 
and the quantum correction in superstring theory. 
We perform a quantitative test of this conjecture at finite temperature, 
by studying the D0-brane matrix quantum mechanics 
and the black zero-brane in type IIA superstring theory. 
We find good agreement, which strongly suggests that the super Yang-Mills theory  does provide us with a nonperturbative formulation of {\it quantum} gravity through 
the gauge/gravity duality. 
\end{abstract}
\maketitle                   





\section{Introduction}
The quantum nature of gravity is one of the most important subjects in theoretical physics. 
It is getting even more important because the Higgs mass measured in LHC suggests that the standard model would be valid all the way up to the Planck scale. 
In this talk we explain the lattice gauge theory can play crucial role in the study of the quantum gravity.  

\begin{figure}
	\begin{center}
		\scalebox{0.3}{\includegraphics{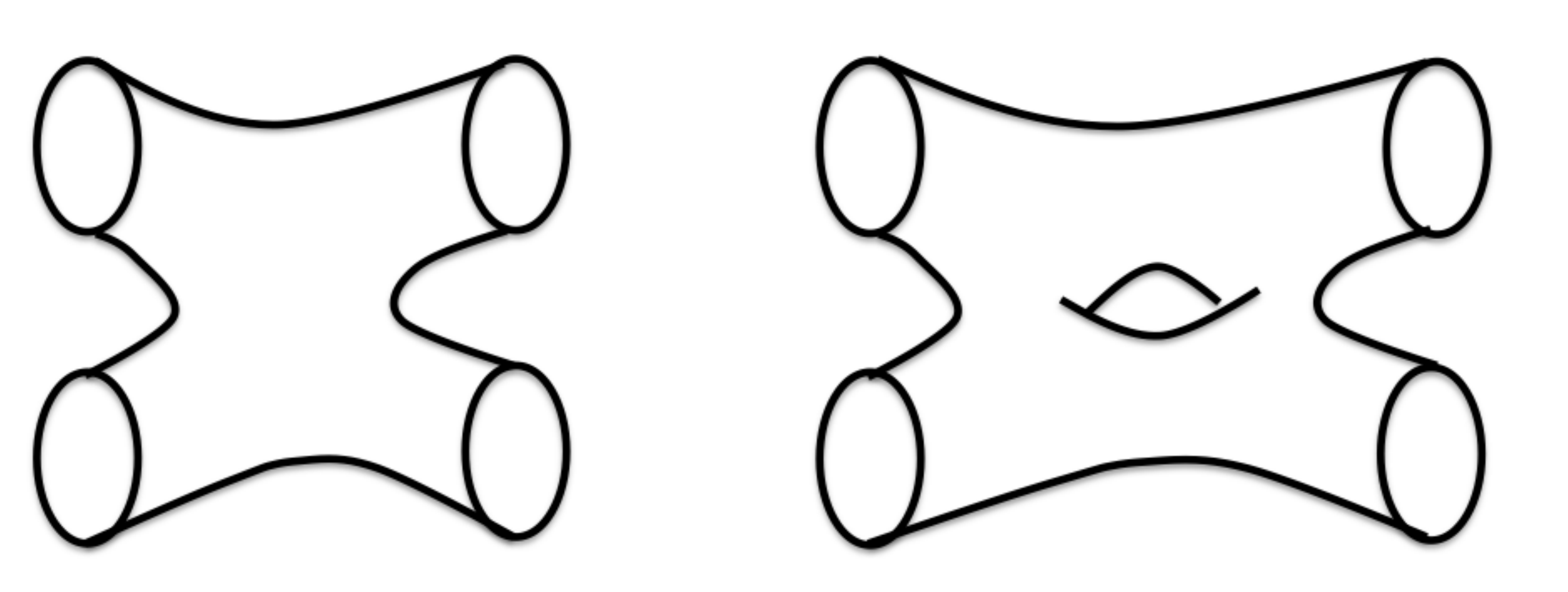}}
	\end{center}
	\caption{Scatterings of two closed strings, tree level (left) and one loop (right). 
	Each closed string loop picks up a factor $g_s^2$. 
	(This figure is taken from \cite{Hanada:2012eg}.)
	}
	\label{fig:string_scattering}
\end{figure}

Superstring theory is a promising candidate 
of the theory of everything, which unifies 
the standard model of particle physics and gravity.
In particular, it provides us with a consistent theory of quantum gravity. 
In Fig.~\ref{fig:string_scattering} scattering of two closed strings is shown. 
Strings propagate from left to right and sweep a two-dimensional surface called `world-sheet.' 
In the left figure, two closed strings interact and merge to a single closed string, and then it splits to two. 
This is the tree level diagram. In the right figure, a loop of closed strings appear as an intermediate state. 
Each closed string loop multiplies a factor $g_s^2$ to the amplitude, where $g_s$ is the string coupling constant; 
a diagram with $g$ closed string loops, which is represented by a genus-$g$ surface, carries a factor $g_s^{2g}$. 
In principle, one can take into account the quantum correction order by order by increasing the genus. 
But it is just a perturbative definition at best. How can one formulate string theory nonperturbatively? 

\begin{figure}
	\begin{center}
		\scalebox{0.3}{\includegraphics{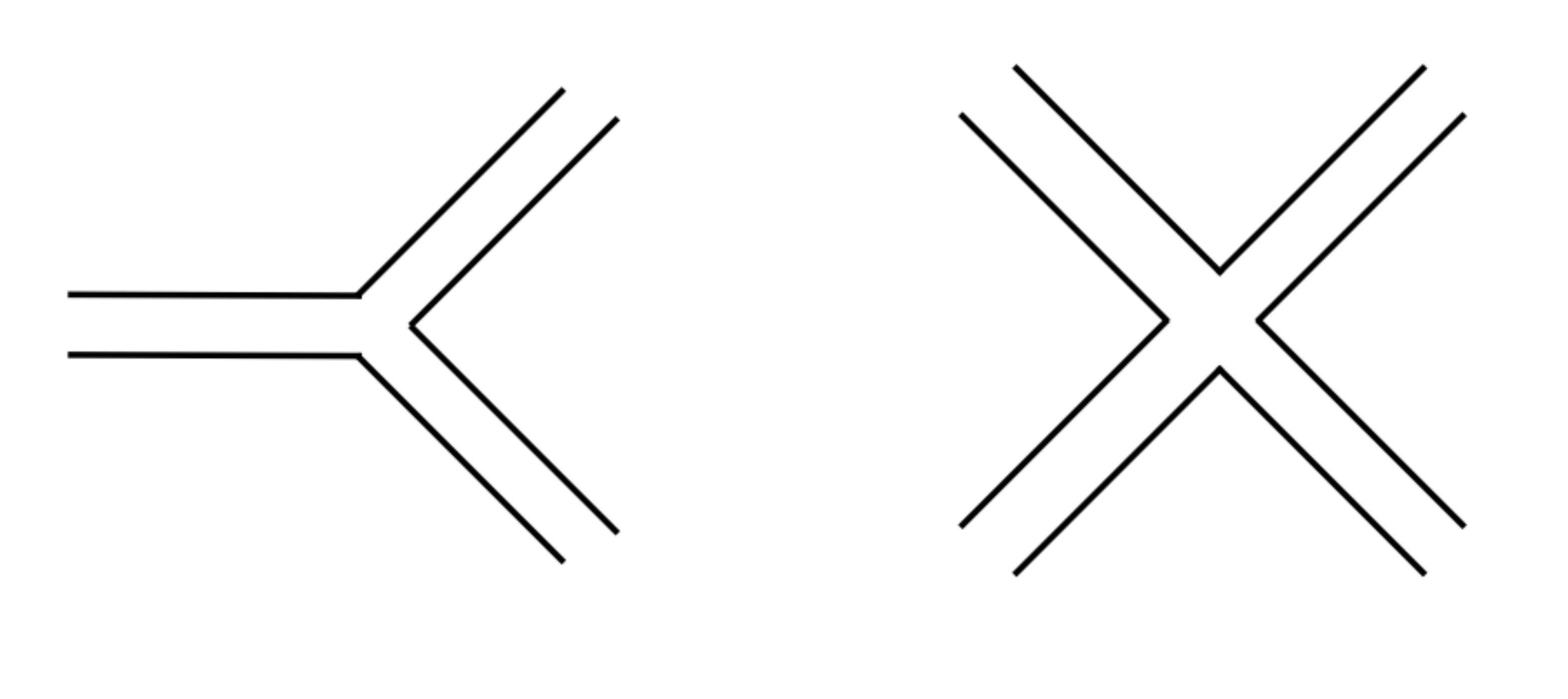}}
	\end{center}
	\caption{Three-point and four-point interaction of the gauge field.}
	\label{fig:interaction_vertices}
\end{figure}

\begin{figure}
	\begin{center}
		\scalebox{0.4}{\includegraphics{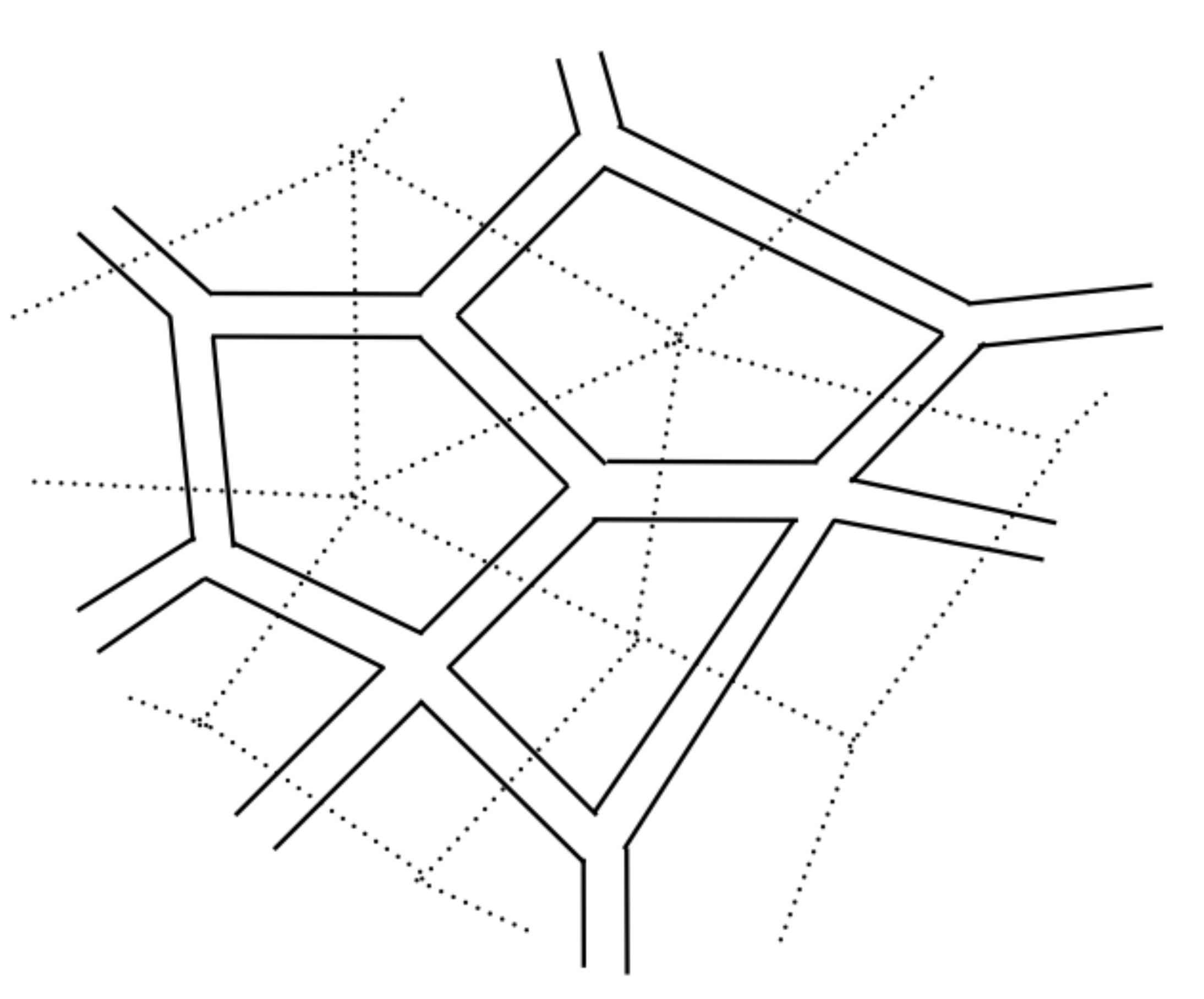}}
	\end{center}
	\caption{Relationship between a Feynman diagram (solid line) and a two-dimensional surface made of 
	triangles and quadrangle (dotted line). Three- and four-point vertices in the Feynman diagram are surrounded by triangles and quadrangles. 
	}
	\label{fig:dynamical_triangulation}
\end{figure}

\begin{figure}
	\begin{center}
		\scalebox{0.3}{\includegraphics{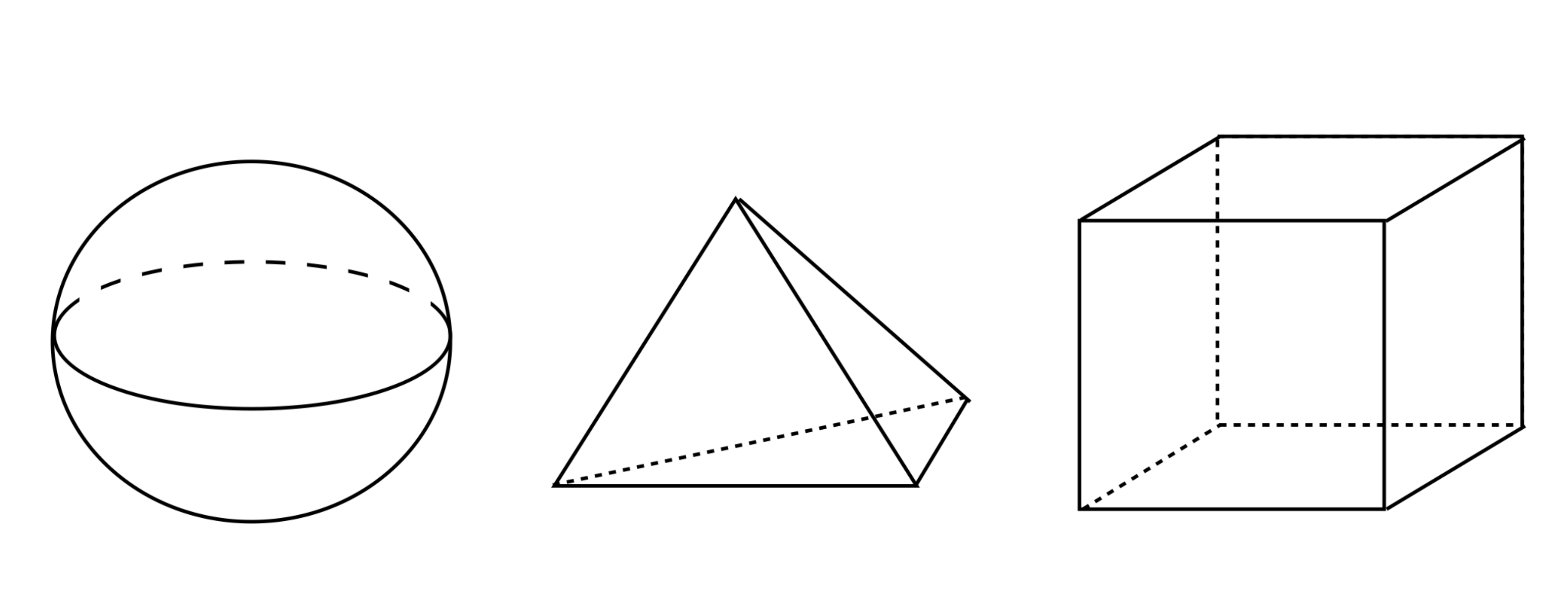}}
	\end{center}
	\caption{Triangulation/quadrangulation of a sphere, $g=0$. 
	${\rm (\#\ triangles)}-{\rm (\#\ sides)}+{\rm (\#\ vertices)}=4-6+4=2$ for the middle and 
	${\rm (\#\ quadrangle)}-{\rm (\#\ sides)}+{\rm (\#\ vertices)}=6-12+8=2$ for the right. }
	\label{fig:sphere}
\end{figure}

\begin{figure}
	\begin{center}
		\scalebox{0.3}{\includegraphics{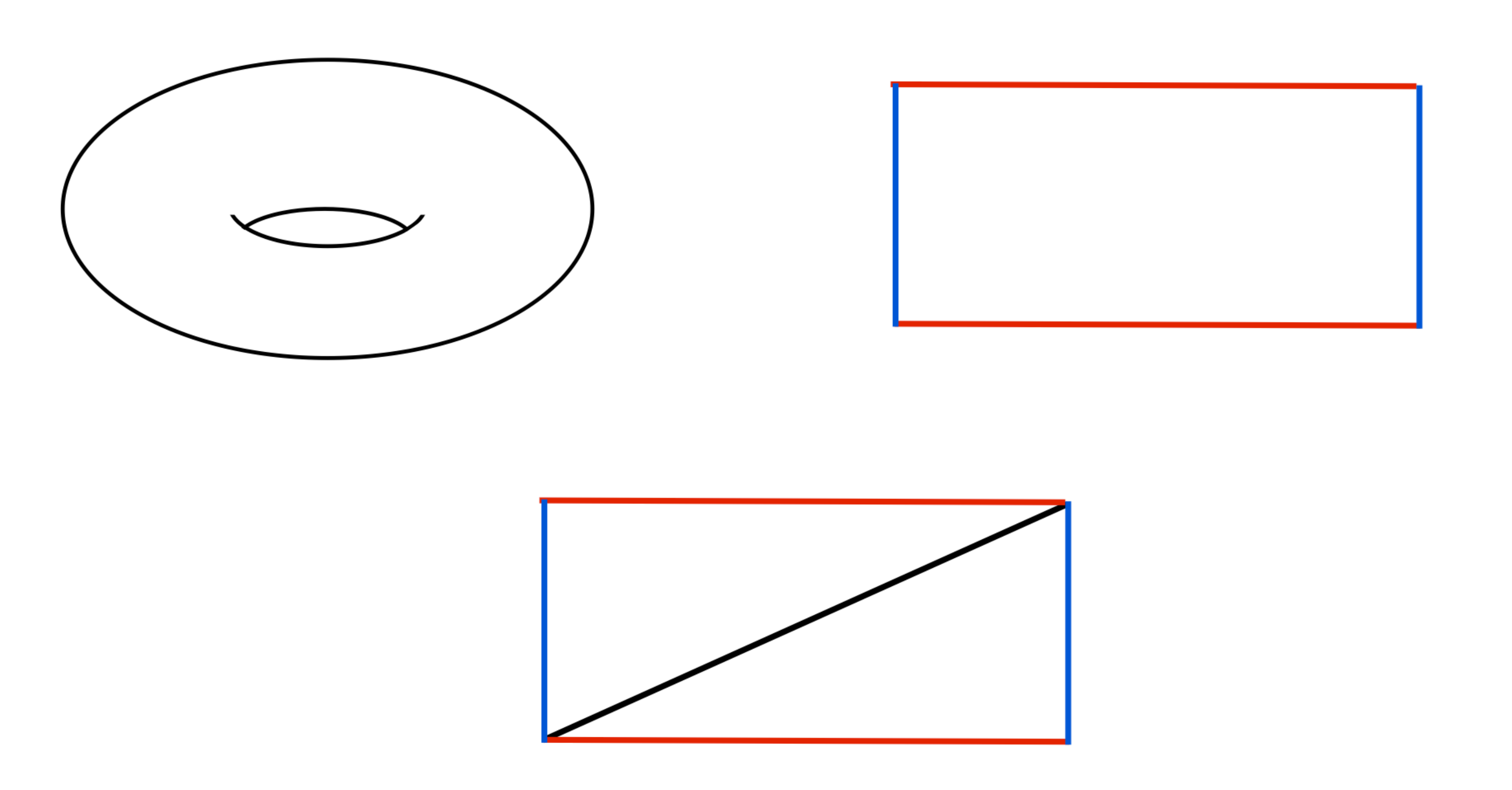}}
	\end{center}
	\caption{Triangulation/quadrangulation of a torus, $g=1$. Red and blue lines are identified, respectively. 
	${\rm (\#\ quadrangle)}-{\rm (\#\ sides)}+{\rm (\#\ vertices)}=1-2+1=0$ for the upper-right and 
	${\rm (\#\ triangles)}-{\rm (\#\ sides)}+{\rm (\#\ vertices)}=2-3+1=0$ for the bottom. }
	\label{fig:torus}
\end{figure}

\begin{figure}
	\begin{center}
		\scalebox{0.3}{\includegraphics{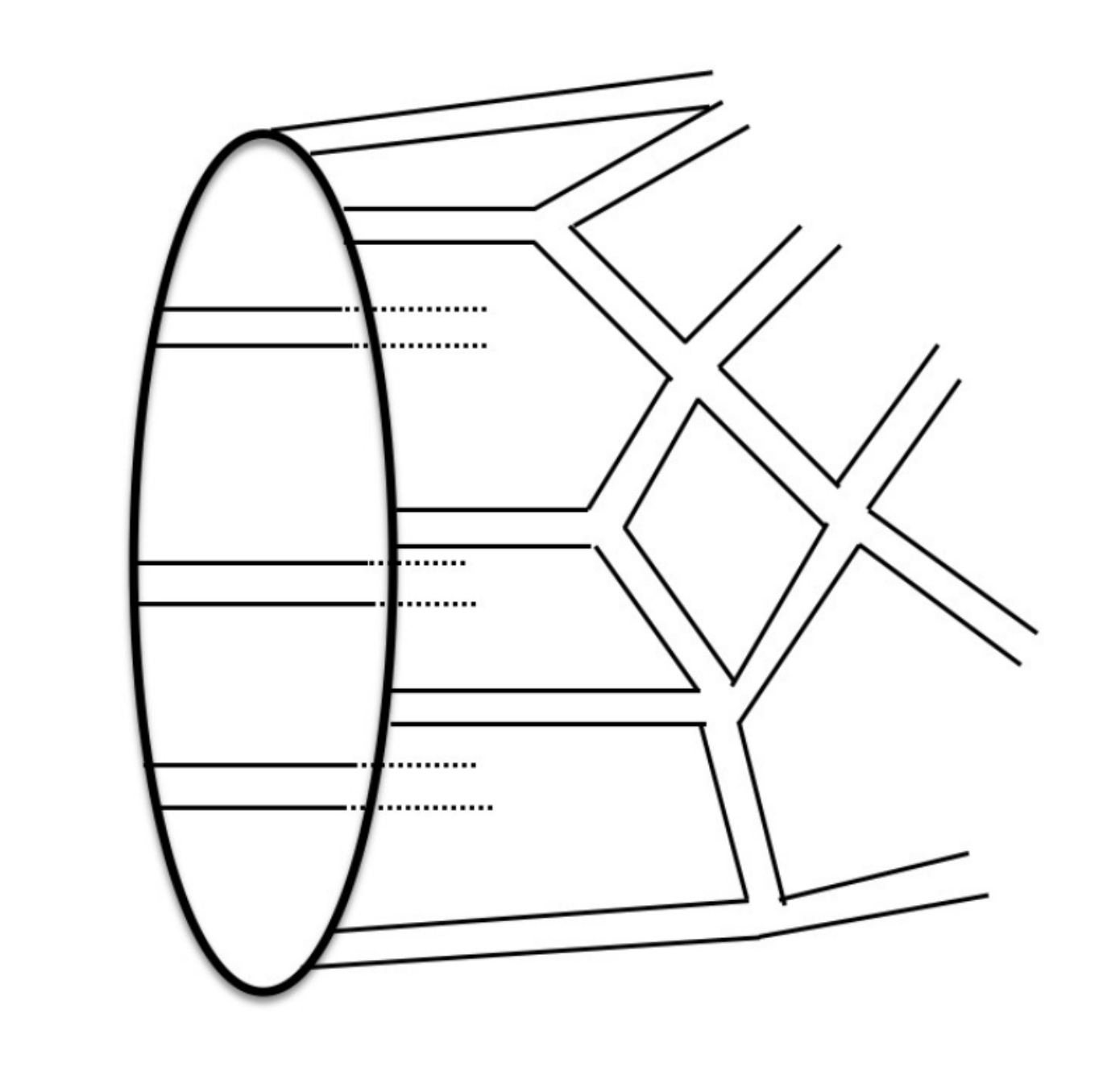}}
	\end{center}
	\caption{The Wilson loop sources many gluons, which form a cylinder-shaped fishnet. This cylinder is identified to the world-sheet of the closed string, 
	and hence the Wilson loop is identified with the creation function of the closed string.}
	\label{fig:wilsonloop}
\end{figure}

In 1970s 't Hooft pointed out a deep relationship between the Yang-Mills theory with the $U(N)$ gauge group and string theory, 
based on the $1/N$ expansion around the 't Hooft large-$N$ limit (i.e. $\lambda\equiv g_{YM}^2N$ fixed). 
Because the gauge field is represented by $N\times N$ matrices, the propagator is represented by two lines, 
which represent the row and column, respectively. 
The interaction is described by three- and four-point vertices (Fig.~\ref{fig:interaction_vertices}). 
A typical example of Feynman diagrams is shown in Fig.~\ref{fig:dynamical_triangulation}.  
It contains the vertices, propagators and closed loops of matrix indices. 
Because the action is written as $(N/4\lambda)\int d^4x TrF_{\mu\nu}^2$, 
each vertex and propagator are associated with $N$ and $N^{-1}$, respectively. 
A closed loop carries a factor $N$ because a sum over the index, which runs from $1$ to $N$, is involved. 
Therefore, each vacuum diagram is proportional to $N^\chi$, where $\chi\equiv {\rm (\#\ vertices)}-{\rm (\#\ propagators)}+{\rm (\#\ index\ loops)}$. 
As a next step, we relate Feynman diagrams to two-dimensional surfaces. Intuitively, a Feynman diagram with many lines 
looks like a dense `fishnet', which eventually becomes indistinguishable from a continuum two-dimensional surface.  
In order to make this intuition mathematically well-defined, let us notice that a Feynman diagram can be translated to a two-dimensional surface made of 
triangles and quadrangles, as shown in Fig.~\ref{fig:dynamical_triangulation}. 
From the figure it is easy to see that 
\begin{eqnarray}
{\rm (\#\ triangles)} &=& {\rm (\#\ three\mathchar`-point\ vertices)}, \nonumber\\
{\rm (\#\ quadrangles)} &=& {\rm (\#\ four\mathchar`-point\ vertices)},\nonumber\\
{\rm (\#\ sides)} &=& {\rm (\#\ propagators)},\nonumber\\
{\rm (\#\ vertices)} &=& {\rm (\#\ index\ loops)}.\\
\end{eqnarray}
Therefore  
\begin{eqnarray}
\chi
=
{\rm (\#\ triangles/quadrangle)}-{\rm (\#\ sides)}+{\rm (\#\ vertices)}
=
2-2g, 
\end{eqnarray}
which is the Euler number of the two-dimensional surface. 
Note that the Euler number depends only on the number of genus $g$,   
as demonstrated in Fig.~\ref{fig:sphere} and Fig.~\ref{fig:torus}.  
Therefore, it is natural to identify the string world-sheet and the coupling constant $g_s$ with the Feynman diagram and $1/N$, 
respectively. In this picture, the closed string corresponds to the Wilson loop, because gluons sourced by the Wilson loop 
form a cylinder-shaped fishnet, which is identified with the world-sheet swiped by the closed string (Fig.~\ref{fig:wilsonloop}). 

If this identification is correct, then the Yang-Mills theory provides us with a nonperturbative formulation of string theory, 
because it is defined without relying on the $1/N$ expansion. (A usual instanton effect $\sim e^{-8\pi^2/g_{YM^2}}$ 
translates into $e^{-N}\sim e^{-1/g_s}$ and hence it is nonperturbative also in the context of string theory.)

This argument is rather formal and it was not clear which specific gauge theory describes string theory. 
More than 20 years after 't Hooft's proposal, Maldacena conjectured that maximally supersymmetric $U(N)$ Yang-Mills theory in $(p+1)$ dimensions 
corresponds to type II superstring theory on the black $p$-brane background \cite{Maldacena:1997re,Itzhaki:1998dd}. 
Here the black $p$-brane is a black hole-like object with $p$-dimensional spatial extension. The black $0$-brane, which we consider below, is just a black hole. 
Therefore, if Maldacena's conjecture is correct, by studying the $1/N$ correction in Yang-Mills theory one can learn about quantum gravity. 

The gauge/gravity duality conjecture 
motivated Kaplan, Katz and Unsal to put super Yang-Mills theories in various dimensions on lattice\cite{Kaplan:2002wv}. 
By now there are various lattice and non-lattice techniques with which one can simulate super Yang-Mills theories in various dimensions. 
In this talk we consider the $(0+1)$-dimensional theory, which is numerically most tractable. We compare the mass of the black hole 
and its counterpart in the gauge theory and find a good agreement at finite values of $N$, which provides us with quantitative evidence 
supporting the Maldacena conjecture at finite-$N$ and quantum gravity level.  
We briefly explain the dual gravity prediction for the gauge theory in Sec.~\ref{sec:duality} and test it by Monte Carlo simulation in Sec.~\ref{sec:test}. 
We discuss implications of our result and future directions in Sec.~\ref{sec:discussion}. 
\section{The gauge/gravity duality and dual gravity prediction}\label{sec:duality}
\begin{figure}[htbp] 
\begin{center}
\scalebox{0.4}{\includegraphics{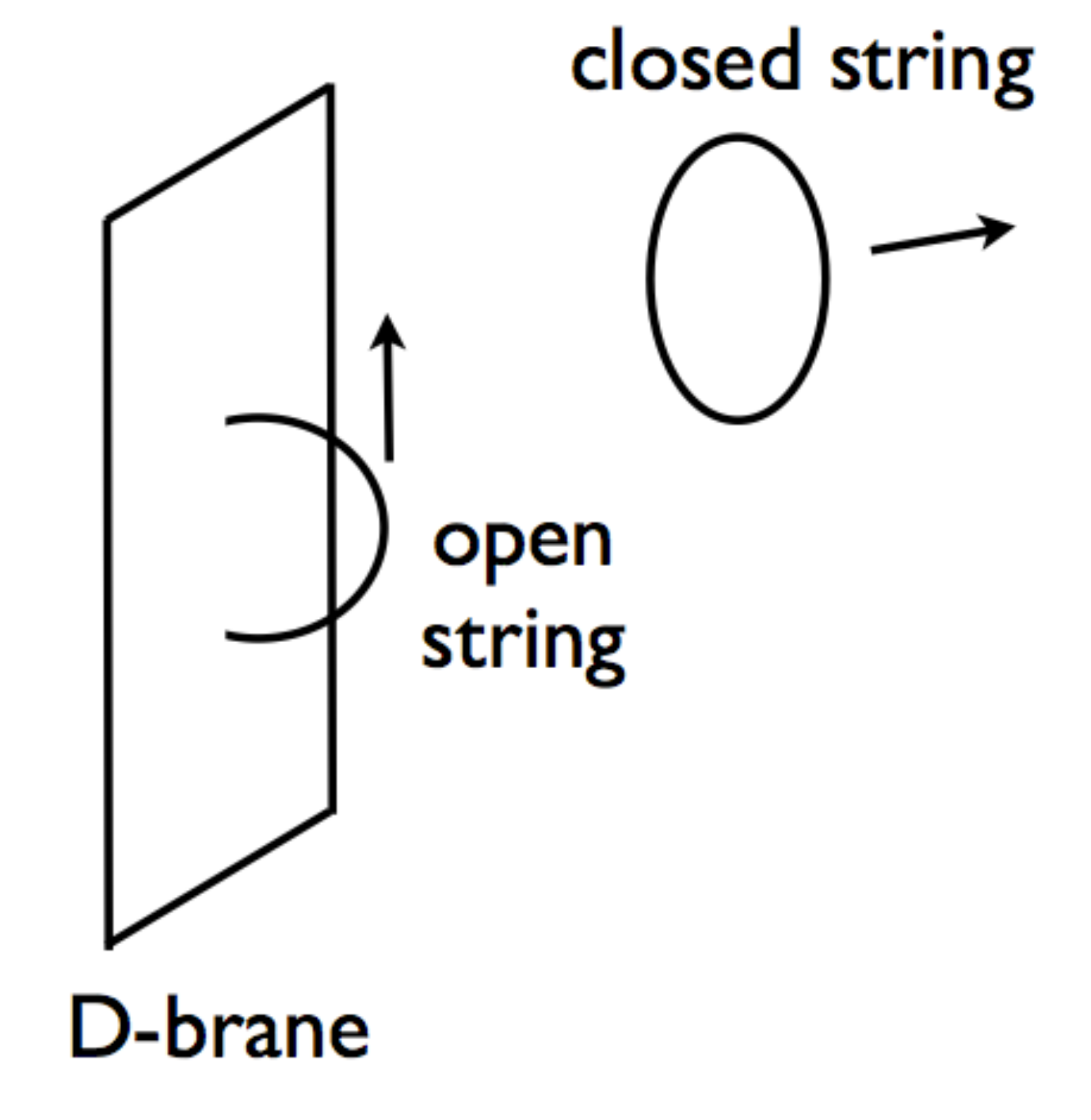}}
\caption{ 
Open strings are attached to D-branes. Closed strings can propagate away from D-branes. (This figure is taken from \cite{Hanada:2012eg}.)
}\label{fig:Dbrane_1}
\end{center} 
\end{figure}

Black $p$-brane consists of a large number of D$p$-branes \cite{Polchinski:1995mt}, 
which are $(p+1)$-dimensional objects on which open strings can have endpoints (Fig.~\ref{fig:Dbrane_1}). 
The low-energy dynamics of $N$ D-branes are described by maximally supersymmetric  $U(N)$ Yang-Mills theory in $(p+1)$ dimensions \cite{Witten:1995im}, 
whose action is given by 
\begin{eqnarray}
S=\frac{1}{g_{YM}^2}\int d^{p+1}x\ Tr\left\{
\frac{1}{4}F_{\mu\nu}^2
+
\frac{1}{2}\left(D_\mu X_i\right)^2
-
\frac{1}{4}
[X_i,X_j]^2
\right\}
+ (fermions), 
\end{eqnarray}
where $\mu,\nu$ run from $1$ to $p+1$ and $X_i$ $(i=1,2,\cdots,9-p)$ are scalar fields. 
The spacetime dimension is $p+1$ because open strings can move only along D-branes. 
Intuitively, $(i,j)$-components of the matrices represent an open string connecting $i$-th and $j$-th D-branes. 
Scalar fields $X_i$ describe the fluctuations of D-branes and open strings to the transverse directions. 

D-branes are massive objects and hence source the gravitational field, or in other words, D-branes emit close strings, 
the spacetime gets curved, and they turn to a black $p$-brane.  
Near the horizon, the metric of the black $p$-brane geometry is given by \cite{Gibbons:1987ps,Itzhaki:1998dd}
\begin{eqnarray}
ds^2
&=&
\alpha'\Biggl\{
\frac{U^{(7-p)/2}}{g_{YM}\sqrt{d_p N}}
\left[
-\left(
1-\frac{U_0^{7-p}}{U^{7-p}}
\right)dt^2
+
dy_{\parallel}^2
\right]
\nonumber\\
& &
\qquad
+
\frac{g_{YM}\sqrt{d_p N}}{U^{(7-p)/2}\left(1-\frac{U_0^{7-p}}{U^{7-p}}\right)}dU^2
+
g_{YM}\sqrt{d_p N}U^{(p-3)/2}d\Omega_{8-p}^2
\Biggl\},
\label{near_extremal_metric} 
\end{eqnarray}
where the Yang-Mills coupling $g_{YM}$ and the size of the gauge group $N$ in the corresponding super Yang-Mills theory are used. 
A constant $\alpha'$ is the square of the string length, 
$(t,y_\parallel)$ represent the $(p+1)$-dimensional extension of the brane,  
$U$ and $\Omega$ are the radial and angular coordinate of the transverse directions, and  
$U_0$ is the place of the horizon.  The Hawking temperature is 
\begin{eqnarray}
T_H=\frac{(7-p)U_0^{(5-p)/2}}{4\pi \sqrt{d_p g_{YM}^2N}},  
\end{eqnarray}
where $d_p=2^{7-2p}\pi^{(9-3p)/2}\Gamma((7-p)/2)$. 
The string coupling constant is given by 
\begin{eqnarray}
g_s = 
(2\pi)^{2-p}g_{YM}^2
\left(
\frac{d_p g_{YM}^2N}{U^{7-p}}
\right)^{\frac{3-p}{4}}. 
\label{string_coupling}
\end{eqnarray}
When $\lambda=g_{YM}^2N$ is fixed, it behaves as $g_s\propto 1/N$, in the same way as in 't Hooft's identification. 
Maldacena conjectured that super Yang-Mills describes full string dynamics around the black brane and 
the $1/N$ correction corresponds to the quantum correction. 
The Hawking temperature and the mass of the black brane are identified with the temperature and the energy $E\equiv -\partial \log Z/\partial \beta$ of the gauge theory. 
In this work we study the case of $p=0$. The gauge theory is quantum mechanics of $N\times N$ matrices, which was originally proposed as the matrix model of M-theory 
\cite{Banks:1996vh}. 
The black hole mass can be calculated by adding stringy corrections to the black $0$-brane geometry shown above. 
The result is \cite{Hyakutake:2013vwa} 
\begin{eqnarray}
\frac{1}{N^2}
  E_{gravity} &= 
(7.41 \, T^{2.8} + a \, T^{4.6}+\cdots) +
  (-5.77 \, T^{0.4} + b \, T^{2.2}+\cdots)\frac{1}{N^2}
+ O\left(\frac{1}{N^4}\right) \ ,
\label{gravity_prediction}
\end{eqnarray}
where $T=\lambda^{-1/3}T_H$ is dimensionless effective temperature and 
$a,b$ are unknown constants. The energy $E_{gravity}$ is also made dimensionless, by multiplying $\lambda^{-1/3}$. 
In the following, we simply set $\lambda=1$ without loss of generality. 
Higher order terms in each power of $1/N$ represent to the $\alpha'$ corrections, 
which appear because strings are not point-like. 
If Maldacena's conjecture is correct, this expression must be reproduced from the matrix quantum mechanics. 

\section{Testing the duality}\label{sec:test}
The $O(N^0)$ terms of the dual gravity prediction (\ref{gravity_prediction}) have been tested previously in 
\cite{Anagnostopoulos:2007fw,Catterall:2008yz,Hanada:2008ez,Kadoh:2012bg}. 
In particular, the $\alpha'$ correction $a \, T^{4.6}$ has been confirmed 
with $a=-5.58(1)$, by looking at $T\gtrsim 0.5$ \cite{Hanada:2008ez}.  
  
In order to study the $1/N$ correction, we must study rather small values of $N$. Here we study $N=3,4$ and $5$. 
(That we have to take $N$ small caused a technical problem which required a proper treatment; see \cite{Hanada:2013}.)
We also have to study low temperature where the $\alpha'$ correction becomes small, because there are too many fitting parameters otherwise. 
For this reason we consider $0.08\le T\le 0.11$. 

\begin{figure}
	\begin{center}
		\scalebox{0.3}{\includegraphics{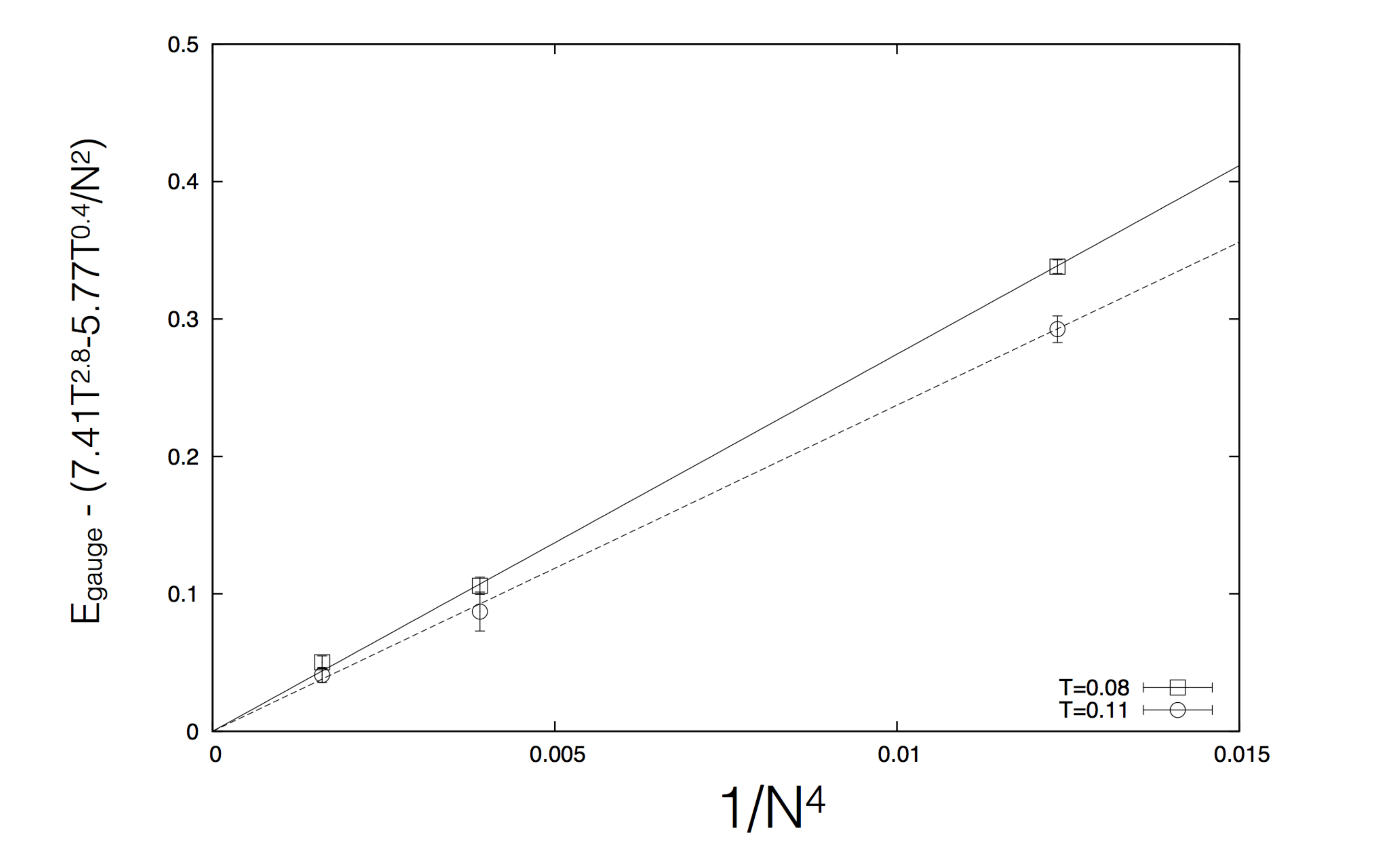}}
	\end{center}
	\caption{$E_{gauge}/N^2 - (7.41T^{2.8}-5.77T^{0.4}/N^2)$. }
	\label{fig:N4_law}
\end{figure}

\begin{figure}
	\begin{center}
	\rotatebox{0}{
		\scalebox{0.7}{\includegraphics{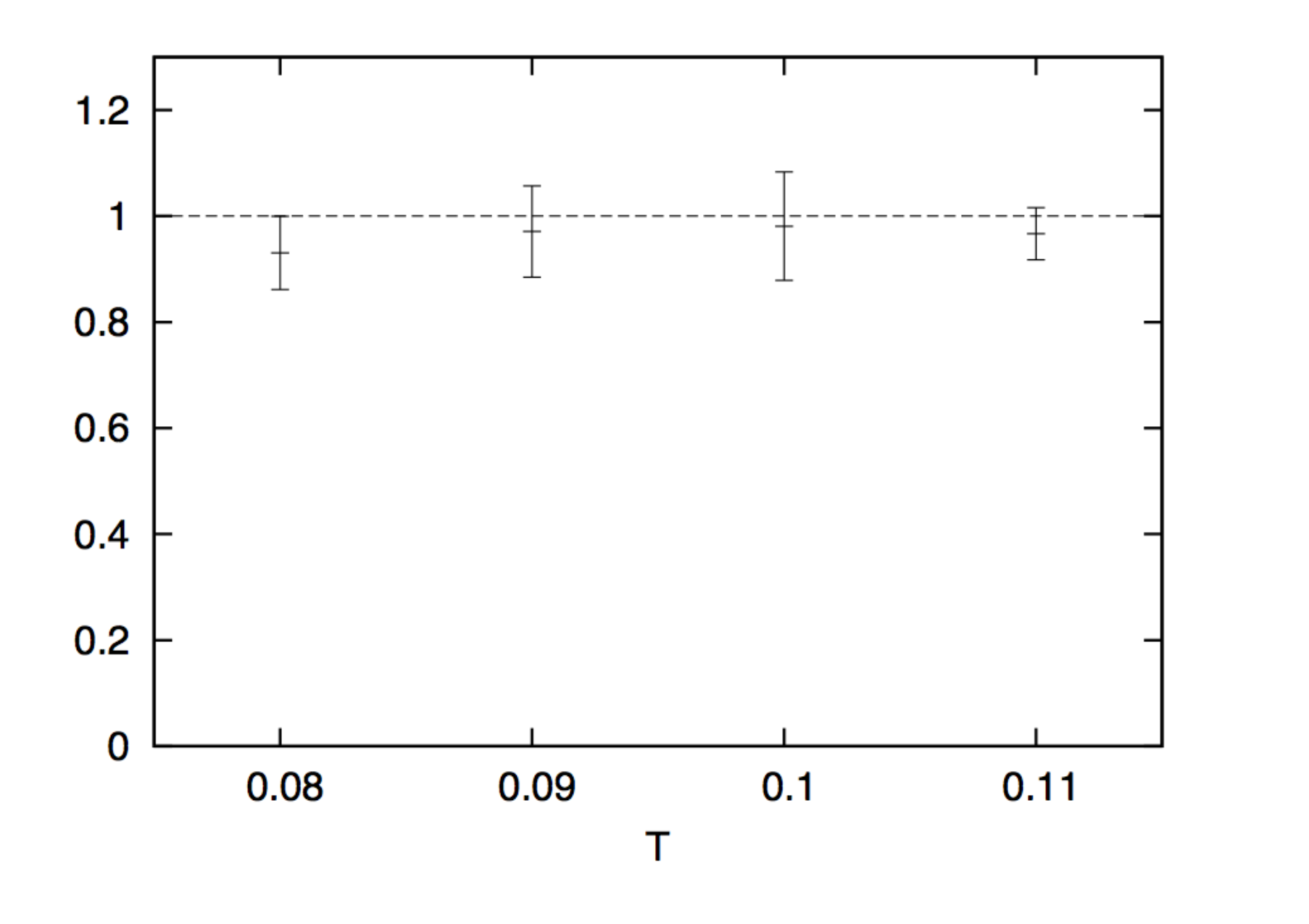}}}
	\end{center}
	\caption{The coefficient of $1/N^2$ divided by $-5.77T^{0.4}$. }
	\label{fig:coeff}
\end{figure}

In Fig.~\ref{fig:N4_law} we plot $E_{gauge}/N^2 - (7.41T^{2.8}-5.77T^{0.4}/N^2)$. Remarkably, this quantity behaves as $1/N^4$.
Therefore the first two terms of the $1/N^2$ expansion agree with the dual gravity prediction, and terms of order $1/N^6$ and higher are small compared to the $1/N^4$ term. 
As a further test, we estimated the coefficient of $1/N^2$ by using a fitting ansatz $E_{gauge}=7.41T^{2.8}+c_1/N^2+c_2/N^4$. 
In Fig.\ref{fig:coeff} we show this coefficient normalized by the dual gravity prediction,  $c_1/(-5.77T^{0.4})$. 
It is reasonably close to $1$. 
This is very strong evidence that the gauge/gravity duality holds at quantum gravity level. 

\section{Discussion}\label{sec:discussion}
We have shown the first quantitative evidence for the gauge/gravity duality at the level of finite $N$ (gauge theory side) 
and quantum gravity (gravity side) beyond kinematics, by studying the finite temperature system in which supersymmetry is broken by the thermal effect. 
We studied mass of the black hole in superstring theory and its counterpart in the Yang-Mills theory, and found good agreement. 
It strongly suggests that, as Maldacena claimed in the first paper on the gauge/gravity duality \cite{Maldacena:1997re}, 
{\it the gauge theory provides us with the nonperturbative formulation of superstring theory}. 
Then, by studying the dynamics of super Yang-Mills theory on computers, we can learn about quantum gravity. 

Furthermore, in \cite{Hanada:2013}, we found that the negative specific heat appears at very low temperature ($T\lesssim 0.10$ for $N=3,4$ and $5$). 
It suggests that the quantum correction drives black hole unstable and makes it evaporate. 
If it can be shown more robustly, then the black hole described through the gauge theory would provides us with a concrete counter-example 
to Hawking's information loss paradox \cite{Hawking:1976ra}.
We expect further numerical study of the Yang-Mills theory will lead us to a better understanding about the information puzzle, especially how the information comes back from the black hole. 
\section*{Acknowledgement}
This talk was based on the collaboration with Y.~Hyakutake, G.~Ishiki and J.~Nishimura \cite{Hanada:2013}. 
I would like to thank them for useful comments on this proceeding.

\end{document}